\documentstyle[12pt,epsf]{article}
\newcommand{\vphi}{\varphi}
\newcommand{\DS}{\displaystyle}

\begin{document}

\title{
Rotating Hairy Black Holes}
\vspace{1.5truecm}
\author{
{\bf Burkhard Kleihaus}\\
Department of Mathematical Physics, University College, Dublin,\\
Belfield, Dublin 4, Ireland\\
{\bf Jutta Kunz}\\
Fachbereich Physik, Universit\"at Oldenburg, Postfach 2503\\
D-26111 Oldenburg, Germany}

\vspace{1.5truecm}

%\date{December 13, 2000}
\date{\today}

\maketitle
\vspace{1.0truecm}

\begin{abstract}
We construct stationary black holes in SU(2) Einstein-Yang-Mills theory,
which carry angular momentum and electric charge.
Possessing non-trivial non-abelian magnetic fields
outside their regular event horizon,
they represent non-perturbative rotating hairy black holes.
\end{abstract}
\vfill
\noindent {Preprint hep-th/0012081} \hfill\break
\vfill\eject

\section{Introduction}

Black holes in Einstein-Maxwell (EM) theory are completely determined
by their mass, their charge and their angular momentum,
i.e.~EM black holes have ``no hair'' \cite{nohair1,nohair2}.
The unique family of stationary asymptotically flat 
EM black hole solutions comprises 
the rotating charged Kerr-Newman (KN) solutions,
the static charged Reissner-Nordstr\o m (RN) solutions,
the rotating Kerr solutions
and the static Schwarzschild solutions.
Besides the ``no-hair'' theorem, Israel's theorem
holds in EM theory, stating that static black holes
are spherically symmetric.

Both theorems cannot be extended to theories with non-abelian 
gauge fields \cite{su2bh,review}.
While the first classical ``hairy'' black hole solutions found are static 
and spherically symmetric \cite{su2bh},
recently also static black hole solutions 
with only axial symmetry have been constructed \cite{kk,map3},
as well as static (perturbative) black hole solutions without rotational
symmetry \cite{ewein}. 

Evidently, also rotating hairy black hole solutions should exist \cite{sudar}.
The construction of such solutions, however, has represented a
difficult challenge. In particular, the usual parametrization
of the stationary axially symmetric metric \cite{wald} has been considered
as possibly too narrow for non-abelian solutions \cite{pert1,gal,heus}, 
and furthermore, even the usual parametrization leads to a highly involved
set of differential equations, to be solved numerically.
Therefore, stationary generalizations of the static spherically symmetric
SU(2) Einstein-Yang-Mills (EYM) black hole solutions \cite{su2bh}
have previously only been considered perturbatively \cite{pert1,pert2}.

Here we construct the first set of non-perturbative
rotating non-abelian black hole solutions.
Considering SU(2) EYM theory as well,
we obtain hairy black hole solutions which are asymptotically flat
and possess a regular event horizon.
Like their perturbative counterparts \cite{pert1},
these hairy black hole solutions carry 
both angular momentum and electric charge.

\section{\bf Embedded Kerr-Newman Black Holes}

We consider the SU(2) EYM action
\begin{equation}
S=\int \left ( \frac{R}{16\pi G} 
- \frac{1}{2} {\rm Tr} (F_{\mu\nu} F^{\mu\nu})
  \right ) \sqrt{-g} d^4x
\ ,   \end{equation}
with $F_{\mu \nu} = 
\partial_\mu A_\nu -\partial_\nu A_\mu + i e \left[A_\mu , A_\nu \right] $,
$A_\mu = 1/2 \tau^a A_\mu^a$,
and $G$ and $e$ are Newton's constant and
the Yang-Mills coupling constant, respectively.
Variation with respect to the metric and the matter fields
leads to the Einstein equations and the field equations,
respectively.

We first note that
Kerr-Newman black holes may be embedded in SU(2) EYM theory \cite{perry}.
For a Kerr-Newman black hole with mass $M$,
angular momentum $L=aM$ and gauge invariant ``total charge'' $Q$, 
$Q^2 = Q^aQ^a + P^a P^a$ \cite{foot0},
the metric in Boyer-Lindquist coordinates is given by
\begin{equation}
ds^2 = -\frac{\Delta}{\rho^2} \left(dt + a \sin^2 \theta d \varphi \right)^2
  + \frac{\sin^2\theta}{\rho^2} \left( a dt + \rho_0^2 d \varphi \right)^2
  + \frac{\rho^2}{\Delta} d \tilde r^2 + \rho^2 d\theta^2
\ , \label{KN0} \end{equation}
with
\begin{equation}
\rho^2 = \tilde r^2+a^2\cos^2\theta \ , \ \ \
\rho^2_0 = \tilde r^2+a^2 \ , \ \ \  \Delta = \tilde r^2-2M \tilde r+a^2+Q^2
\ , \label{KN6} \end{equation}
and the gauge field is given by
\begin{equation}
A_\mu^a dx^\mu =
\frac{ Q^a \tilde r}{\rho^2} \left(
 dt + a \sin^2 \theta d \varphi \right)
 + \frac{ P^a \cos \theta }{\rho^2} \left(
 a dt + \rho_0^2 d \varphi \right)
\ . \label{KN7} \end{equation}

The condition $\Delta(\tilde r_{\rm H})=0$ yields
the regular event horizon of the Kerr-Newman solutions,
$\tilde r_{\rm H} = M + \sqrt{M^2 - (a^2+Q^2)}$.

\section{\bf Ansatz for Hairy Black Holes}

Proving to be adequate,
we employ the usual parametrization of the metric \cite{wald}
to obtain rotating hairy black hole solutions.
In isotropic coordinates the metric reads
\begin{equation}
ds^2 = -fdt^2+\frac{m}{f}\left[dr^2+r^2 d\theta^2\right] 
       +\sin^2\theta r^2 \frac{l}{f}
          \left[d\vphi+\frac{\omega}{r}dt\right]^2 \  
\ , \label{metric} \end{equation}
where $f$, $m$, $l$ and $\omega$ are functions of only $r$ and $\theta$.
This ansatz satisfies the circularity and Frobenius conditions
\cite{circ,wald,nohair2}.
For the gauge potential we choose the ansatz
\begin{equation}
A_\mu dx^\mu 
  =   \Phi dt +A_\vphi (d\vphi+\frac{\omega}{r} dt) 
+\left(\frac{H_1}{r}dr +(1-H_2)d\theta \right)\frac{\tau_\vphi}{2e}
\ , \label{a1} \end{equation}
with
\begin{equation}
A_\vphi=   -\sin\theta\left[H_3 \frac{\tau_r}{2e} 
            +(1-H_4) \frac{\tau_\theta}{2e}\right] \ , \ \ \
\Phi =B_1 \frac{\tau_r}{2e} + B_2 \frac{\tau_\theta}{2e}
\ , \label{a3} \end{equation}
where the symbols $\tau_r$, $\tau_\theta$ and $\tau_\vphi$
denote the dot products of the cartesian vector of Pauli matrices
with the spherical spatial unit vectors, 
(e.g.~$\tau_r= \tau_x \sin \theta \cos \vphi
 + \tau_y \sin \theta \sin \vphi
 + \tau_z \cos \theta$)
and the gauge field functions $H_i$ and $B_i$
depend on only $r$ and $\theta$.

With respect to the residual gauge degree of freedom \cite{kk} 
we choose the gauge condition 
$r\partial_r H_1-2\partial_\theta H_2 =0$.

\section{\bf Boundary conditions}

To obtain stationary axially symmetric black hole solutions
which are asymptotically flat, and possess a regular event horizon,
as well as a finite mass, angular momentum and electric charge,
we need to impose the appropriate set of boundary conditions.

The condition $f(r_{\rm H})=0$ determines 
the event horizon \cite{foot1}.
Regularity of the event horizon then requires the boundary conditions
$f=m=l=0$, $\omega=\omega_{\rm H}={\rm const}$,
$H_1=0$, 
$\partial_r H_2= \partial_r H_3= \partial_r H_4=0$,
$r_{\rm H} B_1+\cos\theta\omega_{\rm H}=0$,
$r_{\rm H} B_2-\sin\theta\omega_{\rm H}=0$
(with the gauge condition $\partial_\theta H_1=0$ 
taken into account).

The boundary conditions at infinity are
$f=m=l=1$, $\omega=0$,
$H_1 =H_3 =  0$, $H_2=H_4 = \pm 1$, $B_1 = B_2 = 0$.
The boundary conditions on the symmetry axis ($\theta=0$) are
$\partial_\theta f = \partial_\theta l = 
\partial_\theta m = \partial_\theta \omega = 0$,
$H_1 =  H_3 = B_2=0$,
$\partial_\theta H_2 = \partial_\theta H_4 = \partial_\theta B_1=0$,
and agree with
the boundary conditions on the $\theta=\pi/2$-axis,
except for $B_1 = 0$, $\partial_\theta B_2 = 0$.

\section{\bf Properties of the Solutions}

The global charges of the black hole solutions
are determined from their asymptotic behaviour.
In particular, expansion at infinity yields \cite{islam}
\begin{equation} 
f \longrightarrow 1-\frac{f_\infty}{r}, \ \ 
\omega \longrightarrow \frac{\omega_\infty}{r^2}, \ \ 
B_1  \longrightarrow \frac{B_\infty \cos\theta}{r}, \ \ 
B_2  \longrightarrow \frac{B_\infty \sin\theta}{r},
\label{asym} \end{equation}
determining
the mass ${\DS M= \lim_{r\rightarrow\infty}r^2\partial_r f}$,
the angular momentum
${\DS J = \frac{1}{2}\lim_{r\rightarrow\infty}r^2\omega}$
and the electric charge $Q=B_\infty/e$,
which we read off in a gauge where
${\DS \Phi \longrightarrow \frac{B_\infty}{r}\frac{\tau_z}{2e}}$.

Of interest are also the properties of the horizon.
The surface gravity is obtained from \cite{wald}
\begin{equation}
\kappa^2_{\rm sg} = -1/4 (D_\mu \chi_\nu)(D^\mu \chi^\nu) 
\ , \label{sgwald} \end{equation}
where the Killing vector
$\chi = \xi -(\omega_{\rm H}/x_{\rm H}) \eta$ ($\xi=\partial_t$,
$\eta=\partial_\vphi$)
is orthogonal to and null on the horizon.
Expansion near the horizon 
in $\delta = (r-r_{\rm H})/r_{\rm H}$ 
yields to lowest order $f=\delta^2 f_2(\theta)$,
$m=\delta^2 m_2(\theta)$ \cite{long},
and the surface gravity 
\begin{equation}
\kappa_{\rm sg} = \frac{f_2(\theta)}{r_{\rm H} \sqrt{m_2(\theta)}} 
\  \label{temp} \end{equation}
is indeed constant on the horizon \cite{long},
as required by the zeroth law of black hole mechanics.

We further consider the area $A$ of the
black hole horizon, defining the area parameter $r_\Delta$ via
$A = 4 \pi r_\Delta^2$,
and the deformation of the horizon, quantified by the ratio $L_e/L_p$
of the circumferences along the equator and the poles.
We note, 
that the Kretschmann scalar is finite at the horizon \cite{long}.

\section{Numerical Results}

We solve the set of ten coupled non-linear
elliptic partial differential equations numerically,
subject to the above boundary conditions,
employing compactified dimensionless coordinates,
$\bar x = 1-(x_{\rm H}/x)$ with $x=(e/\sqrt{4\pi G}) r$.

The solutions depend on one discrete parameter,
the node number $n$ of the gauge field,
and on two continuous parameters,
the isotropic horizon radius $x_{\rm H}$ and the value of
the metric function $\omega$ at the horizon, $\omega_{\rm H}$,
where $\omega_{\rm H}/x_{\rm H}$ represents the rotational velocity
of the horizon. 

As initial guess we employ the static spherically symmetric 
SU(2) EYM black hole solution
with horizon radius $x_{\rm H}$ and one node,
corresponding to $\omega_{\rm H} =0$ \cite{foot4}.
Increasing $\omega_{\rm H}$ leads to rotating black hole solutions
with non-trivial functions $\omega$, $B_1$, $B_2$, $H_1$, $H_3$,
whose mass $M$, electric charge $Q$ and angular momentum $J$
are determined from their asymptotic behaviour
(see eq.~(\ref{asym})).

As we increase $\omega_{\rm H}$ from zero, while keeping $x_{\rm H}$ fixed, 
a first branch of solutions forms, the lower branch.
This branch extends up to a maximal value of $\omega_{\rm H}$,
which depends on $x_{\rm H}$.
There a second branch, the upper branch, bends backwards towards
$\omega_{\rm H}=0$.
Along both branches mass,
electric charge and angular momentum continuously increase, as seen in Fig.~1,
where the dimensionless mass $\mu =(e/\sqrt{4\pi G})GM$,
the electric charge $Q$ and the angular momentum per unit mass $a=J/\mu$
are shown as functions of $\omega_{\rm H}$ for $x_{\rm H}=1$.

The presence of two branches is no surprise. Indeed, also
the KN and Kerr solutions exhibit two branches, when considered
as functions of $\omega_{\rm H}$
for fixed isotropic horizon radius $x_{\rm H}$.

Whereas both mass $\mu$ and angular momentum per unit mass $a$ 
of the non-abelian solutions increase strongly 
along the upper branch, diverging with $\omega_{\rm H}^{-1}$ in the 
limit $\omega_{\rm H} \rightarrow 0$,
their electric charge $Q$ remains small. 
For comparison, we therefore show in Fig.~1 also $\mu$ and $a$ 
of the corresponding Kerr solutions ($Q=0$) for $x_{\rm H}=1$,
which satisfy $\frac{\omega_{\rm H}}{x_{\rm H}}
=\frac{\sqrt{\mu^2-4 x_{\rm H}^2}} {2\mu(\mu+2x_{\rm H})}$.
The Kerr solutions exist up to a slightly higher value of $\omega_{\rm H}$.
Along the upper branch, 
the non-abelian fields become less important,
and the solutions tend towards extremal Kerr (KN) solutions
in the limit $\omega_{\rm H} \rightarrow 0$.

In Fig.~2 we show the surface gravity $\kappa_{\rm sg}$ for
$x_{\rm H} =1$ as a function of $\omega_{\rm H}$, 
as well as the area parameter $x_\Delta$ and the deformation of the horizon,
quantified by $L_e/L_p$.
As a typical example of a rotating hairy black hole solution, 
we show in Fig.~3 the energy density $-T_0^0$ (of the gauge fields)
for isotropic horizon radius $x_{\rm H}=1$ and $\omega_{\rm H} =0.05$.
Properties of this solution are shown in Table~1,
and compared to the corresponding Kerr values.

In Fig.~4 we show the mass $\mu$ of the non-abelian black hole solutions 
as a function of the isotropic horizon radius $x_{\rm H}$
for several fixed values of $\omega_{\rm H}$.
For a given value of $\omega_{\rm H}$ there is 
a minimum value of the horizon radius $x_{\rm H}$.
In particular, the limit $x_{\rm H} \rightarrow 0$
is only reached for $\omega_{\rm H} \rightarrow 0$.
Thus we do not obtain globally regular rotating solutions in
the limit $x_{\rm H} \rightarrow 0$ \cite{pert1}. This is to be expected,
since globally regular rotating solutions should satisfy
a different set of boundary conditions at infinity \cite{pert2}.

For comparison, we have included in Fig.~4
the mass $\mu$ of the corresponding Kerr black hole solutions.
For a fixed value of $\omega_{\rm H}$,
the Kerr solutions form two straight lines, extending from the origin.
The non-abelian solutions tend toward these lines
for large values of the horizon radius.

To compare with the perturbative calculations,
where linear rotational excitations of the static EYM black holes were studied,
we consider the limit $\omega_{\rm H} \rightarrow 0$ along the lower branch.
In the perturbative calculations $Q \propto J$ \cite{pert1},
and the ratio $Q/J$ depends only on the horizon radius.
The non-perturbative calculations show good agreement
with the non-perturbative results for small values of
$\omega_{\rm H}$ (on the lower branch)
and large values of the horizon radius.
For small values of the horizon radius, significant
deviations arise.

Further details of these rotating non-abelian black hole solutions
as well as the presentation of the rotating higher node solutions
will be given elsewhere \cite{long}.

Beside these non-abelian stationary charged black hole solutions 
with finite angular momentum $J$ and finite electric charge $Q$,
the perturbative studies \cite{pert2} have revealed two more types 
of stationary non-abelian black hole solutions, 
namely rotating black hole solutions which are uncharged 
($J>0$, $Q = 0$),
and non-static black hole solutions, which
have vanishing angular momentum ($J=0$, $Q\ne 0$).
Both types satisfy a different set of boundary conditions at infinity.
Construction of their non-perturbative counterparts
remains open,
as well as the possible existence of
rapidly rotating branches of non-abelian black holes solutions,
not connected to the static solutions.

By including dilaton and axion fields, finally,
further interesting rotating hairy black holes should be generated,
representing new solutions of the low energy effective action
of string theory.

\newpage

\begin{center}
\begin{tabular}{|c|c|c|c|c|c|c|}
\hline
 & \multicolumn{2}{c|}{$\omega_H= 0.01$} 
 & \multicolumn{2}{c|}{$\omega_H= 0.02$} 
 & \multicolumn{2}{c|}{$\omega_H= 0.05$} \\
 \cline{2-7}
 & EYM & Kerr &   EYM & Kerr &  EYM & Kerr \\
 \hline
$\mu$ & $2.23$ & $2.01$ & $2.25$ & $2.03$ & $2.46$ &  $2.20$\\
Q  & $0.233\times 10^{-2}$ & $0$ & $0.474\times 10^{-2}$ & $0$ & $1.35\times 10^{-2}$ & $0$ \\
a  & $0.18$ & $0.16$ & $0.365$ & $0.321$ & $1.051$ & $0.927$ \\
$\kappa$ & $0.112$ & $0.124$ & $0.110$ & $0.122$ & $0.095$ & $0.108$ \\
$x_\Delta$ & $4.23$ & $4.01$ & $4.26$ & $4.04$ & $4.58$ & $4.31$ \\
$L_e/L_p$ & $1.0013$ & $1.0012$ & $1.0055$ & $1.0049$ & $1.0412$ &  $1.0361$\\
\hline
\end{tabular}
\end{center}

Table 1

The dimensionless mass $\mu$, the charge $Q$,
the angular momentum per unit mass $a=L/\mu$,
the surface gravity $\kappa_{\rm sg}$,
the area parameter $x_\Delta$ and the ratio of circumferences $L_e/L_p$
of the rotating non-abelian black hole solutions are shown
for the values $\omega_{\rm H}=0.01$, 0.02 and 0.05
and horizon radius $x_{\rm H}=1$.
For comparison the values of the corresponding Kerr solutions ($Q=0$)
are also shown.

\newpage

\begin{figure}\centering\epsfysize=8cm
\mbox{\epsffile{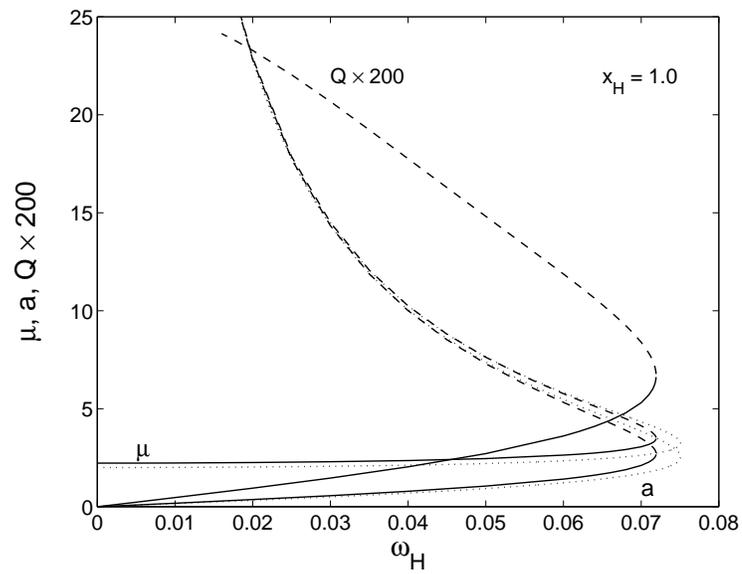}}
\caption{
The dimensionless mass $\mu$ 
of the rotating non-abelian black hole solutions is shown
on the lower branch (solid) and upper branch (dashed)
as a function of the parameter $\omega_{\rm H}$
for the horizon radius $x_{\rm H}=1$.
Also shown are the angular momentum per unit mass $a=L/\mu$
and the charge $Q$.
For comparison $\mu$ and $a$ of the Kerr solution ($Q=0$)
for $x_{\rm H}=1$ are also shown (dotted).
}
\end{figure}

\newpage

\begin{figure}\centering\epsfysize=8cm
\mbox{\epsffile{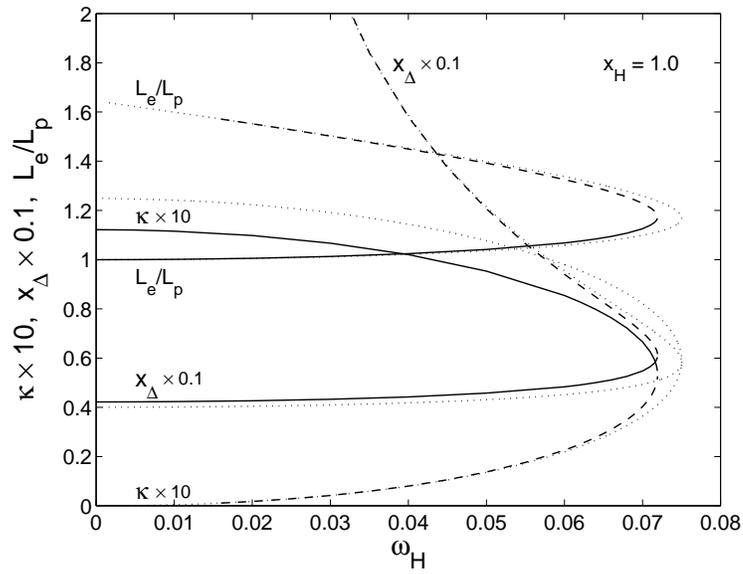}}
\caption{
The surface gravity $\kappa_{\rm sg}$ 
of the rotating non-abelian black hole solutions is shown
on the lower branch (solid) and upper branch (dashed)
as a function of the parameter $\omega_{\rm H}$
for the horizon radius $x_{\rm H}=1$.
Also shown are the area parameter $x_\Delta$ and the deformation,
as quantified by the ratio of circumferences $L_e/L_p$.
For comparison $\kappa_{\rm sg}$, 
$x_\Delta$ and $L_e/L_p$ of the Kerr solution ($Q=0$)
for $x_{\rm H}=1$ are also shown (dotted).
}
\end{figure}

\newpage

\begin{figure}\centering\epsfysize=8cm
\mbox{\epsffile{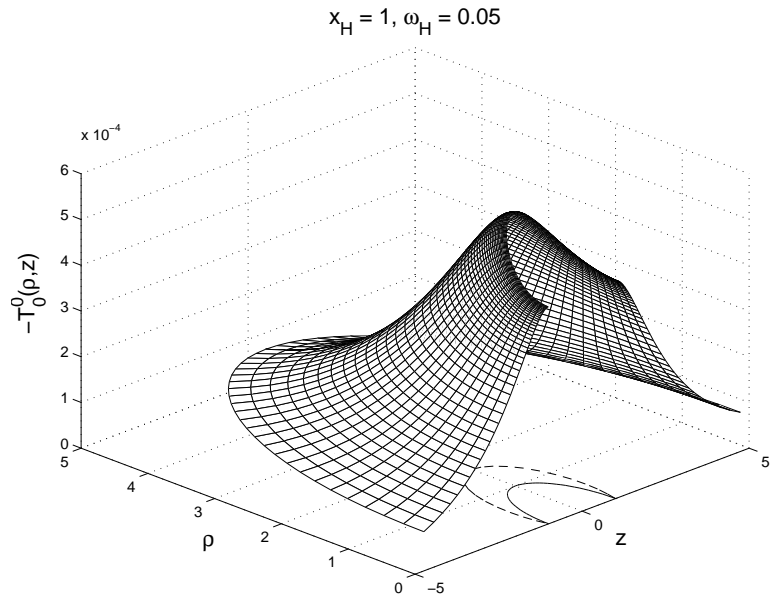}}
\caption{
The energy density $-T_0^0$ 
of the rotating non-abelian black hole solutions is shown
on the lower branch 
for $\omega_{\rm H}=0.05$ and horizon radius $x_{\rm H}=1$.
Also shown are the event horizon (solid) and the static limit (dashed),
enclosing the ergosphere.
}
\end{figure}

 \newpage

\begin{figure}\centering\epsfysize=8cm
\mbox{\epsffile{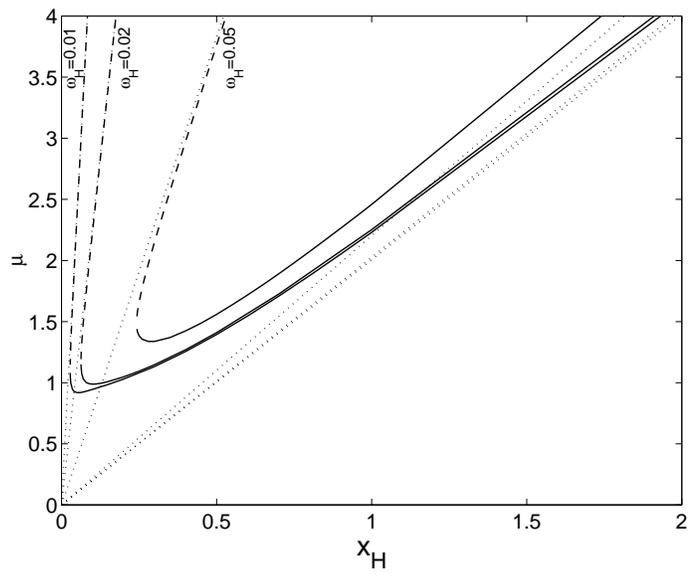}}
\caption{
The dimensionless mass $\mu$
of the rotating non-abelian black hole solutions is shown
on the lower branch (solid) and upper branch (dashed)
as a function of the horizon radius $x_{\rm H}$
for the values of the parameter $\omega_{\rm H}=0.01$, 0.02, 0.05.
For comparison for the same values, the mass of the Kerr solution ($Q=0$)
is also shown (dotted).
}
\end{figure}


\begin{thebibliography}{000}
\bibitem{foot0}
In $Q^2 = Q^aQ^a + P^a P^a$,
$Q^a$ and $P^a$ are constant vectors in the Lie algebra,
considered the Yang-Mills analogue of the electric and
magnetic charge, respectively \cite{perry}.
\bibitem{foot1}
Kerr-Newman solutions have
$f= \frac{\rho^2 \Delta}{\rho_0^4 - a^2 \sin^2 \theta \Delta}$,
$\frac{\omega}{r} =
 \frac{a(2M \tilde r-Q^2)}{\rho_0^4 - a^2 \sin^2 \theta \Delta}$,
$\frac{r^2 l}{f} =
 \rho_0^2 +a^2\sin^2\theta
              \frac{2M \tilde r-Q^2}{\rho^2}$,
where the isotropic coordinate $r$ is related to the Boyer-Lindquist
coordinate $\tilde r$ by
$r(\tilde r)= \frac{1}{2}\left[\sqrt{\tilde r^2-2M \tilde r+(a^2+Q^2)}
 +(\tilde r-M)\right] $,
yielding the isotropic event horizon radius
$r_{\rm H} = 1/2 \sqrt{M^2 - (a^2+Q^2)}$.
\end{thebibliography}
\end{document}